\documentclass[12pt]{iopart}
\usepackage{cite}  
\usepackage{graphicx}
\usepackage{datetime2}
\usepackage{hyperref}
\usepackage{color}
\usepackage{xr}
\usepackage{amssymb} 
\usepackage{relsize}
\begin{document}

\title{Suppression of hyperuniformity in hydrodynamic scalar active field theories}

\author{Nadia Bihari Padhan \& Axel Voigt}

\address{Institute of Scientific Computing, TU Dresden, 01069 Dresden, Germany}
\ead{nadia$\_$bihari.padhan@tu-dresden.de}
\vspace{10pt}

\date{24 November 2024}

\begin{abstract}
The coarsening dynamics at late times in phase-separating systems lead to universally hyperuniform patterns. This is well known for scalar field theories, such as the Cahn-Hilliard equation, but has also been shown for dry scalar active field theories. We demonstrate the role of hydrodynamic interactions in influencing hyperuniformity in a wet active system described by active model H. Our direct numerical simulations reveal that, while (passive) model H shows hyperuniformity in the coarsening regime, the interplay of activity and hydrodynamic interactions suppresses hyperuniformity in active model H, especially when the activity generates contractile stress in the fluid.
\end{abstract}

\section{Introduction}
\label{sec:intro}
Hyperuniformy is a form of hidden order at large scales where density fluctuations are anomalously suppressed \cite{Tor18}. While originally described in terms of density fluctuations in a point pattern the concept of hyperuniformity can be generalized to continuous fields \cite{torquato2016hyperuniformity}. In the past decade, an increasing number of non-equilibrium systems were found to have dynamic hyperuniform states \cite{lei2024non}, among them phase separating systems via spinodal decomposition described by model B or the Cahn-Hilliard equation, which consider a continuous scalar order parameter $\phi(\mathbf r, t)$. The coarsening dynamics at late times in such systems lead to universal hyperuniform patterns. While well known for these passive systems, such states have also recently been observed in dry active matter systems \cite{zheng2024universal, de2024hyperuniformity}, that are described by active field theories, such as active model B and active model B+~\cite{wittkowski2014scalar, fausti2021capillary}, respectively. These models have been developed to describe motility induced phase separation (MIPS) in dry active systems by adding symmetry-breaking terms to the chemical potential, which cannot be derived from the free energy functional. The hyperuniform structures in such scalar fields are characterized by the spectral density $\psi(k)$, where $\psi(k) \rightarrow 0$ according to a power-law behaviour $\psi(k) \sim k^{\alpha}$, with $\alpha > 0$, as the wave number $k \equiv |\mathbf k| \rightarrow 0$~\cite{torquato2016hyperuniformity}. Numerical studies for (passive) model B \cite{ma2017random} and active model B and active model B+ \cite{zheng2024universal} found $\alpha \simeq 4$. A theoretical foundation for $\alpha = 4$ in all of these models was recently provided \cite{de2024hyperuniformity}, extending arguments for the Cahn-Hilliard equation in \cite{Tomita1991}. The essential idea behind the argumentation is the fact that the system is isotropic at scales much larger than the coarsening scale $L$. For this characteristic length scale a power-law growth $L(t) \sim t^{1/3}$ is known for the Cahn-Hilliard equation \cite{lifshitz1961kinetics}. This scaling behaviour translates into the scaling of the spectral density and even under the influence of noise universal scaling at late times with $\alpha = 4$ emerges. While the power-law for $L(t)$ for active field theories is less clear \cite{wittkowski2014scalar}, numerically observed exponents are close to $1/3$ and thus allow for the same argument \cite{de2024hyperuniformity}.

Surprisingly the role of hydrodynamic effects in generating hyperuniform structures in phase-separating systems, both passive and active, remains almost unexplored. However, the interplay between hydrodynamics and activity has been shown to be important for yielding hyperuniform states in active fluids~\cite{backofen2024nonequilibrium,zhang2022hyperuniform}. Phase separation in passive and active fluids with hydrodynamic interactions is governed by model H or the Cahn-Hilliard-Navier-Stokes equations \cite{hohenberg1977theory,anderson1998diffuse,kim2004conservative,giorgini2019uniqueness} and active model H \cite{tiribocchi2015active}, respectively, where the order parameter $\phi(\mathbf r, t)$ is coupled to the velocity field $\mathbf{u}(\mathbf r, t)$ through passive and active stresses in the Navier-Stokes equations. The coarsening mechanism in model H is dominated by hydrodynamic interactions and leads to faster growth in the coarsening length $L(t)$. Specifically, $L(t) \sim t^1$ in the viscous regime~\cite{siggia1979late} and $L(t) \sim t^{2/3}$ in the inertial regime~\cite{furukawa1985effect,kendon19993d}, compared to the slower growth in the diffusion dominated (passive) model B, where $L(t) \sim t^{1/3}$. In active model H, the system undergoes complete phase separation when the active stress is extensile, while a contractile active stress arrests the phase separation, resulting in a non-equilibrium statistically steady state~\cite{tiribocchi2015active}. Particularly in the contractile case, the system exhibits intriguing behaviours such as active turbulence, where the kinetic energy spectrum shows a power-law with scaling exponent similar to the inverse-cascade exponent observed in two-dimensional inertial turbulence~\cite{padhan2024novel}. 

In this work, we demonstrate that hyperuniform structures are robustly maintained in model H, with $\psi(k) \sim k^4$ in the small wavenumber limit, akin to the behaviour in (passive) model B. However, our findings reveal that this is not the case with active model H for contractile stress, where the interplay between the activity and hydrodynamic interactions suppresses the hyperuniform structures in non-equilibrium steady states. We observe homogenization of the fluctuations in $\phi(\mathbf r, t)$, where $\psi(k) \sim \textit{constant}$ for the limit $k \rightarrow 0$. Through a scale-by-scale analysis of the order parameter spectrum equation, we observe that the spectral energy transfer associated with advection term drives the system towards either hyperuniform or non-hyperuniform states. Our findings further reveal that the energy flux $\Pi(k)$ indicates an inverse cascade of structural energy in cases that show complete phase separation. In contrast, for active contractile fluids, this inverse cascade is absent, and the flux instead suggests a forward cascade that prevents hyperuniformity by sustaining fluctuations across multiple length scales. 

The rest of the paper is organised as follows: in Section 2 we give details of the mathematical model and the direct numerical simulations; in Section 3 we describe our results and in Section 4 we draw conclusions.

\section{Model and Simulations}
The dynamics of the order parameter $\phi(\mathbf r, t)$ and the velocity field $\mathbf u (\mathbf r, t)$ is governed by the following coupled hydrodynamic equations:
\begin{eqnarray}
    \partial_t \phi + (\mathbf u \cdot \nabla) \phi &=& M \nabla^2 \mu \label{eq:ch}\\
    \partial_t \mathbf u + (\mathbf u \cdot \nabla) \mathbf u &=& -\nabla p + \nu \nabla^2 \mathbf u + \nabla \cdot \Sigma, \;\;\; \nabla \cdot \mathbf u = 0,\label{eq:ns}
\end{eqnarray}
where, $M$ is a (constant) mobility, $\nu$ is the kinematic viscosity, and $p(\mathbf r, t)$ is the pressure. The chemical potential $\mu(\mathbf r, t)$ can be expressed as a functional derivative of a Landau-Ginzburg-type free energy functional $\mathcal F$ with respect to the order parameter $\phi$~\cite{cates2018theories}:
\begin{eqnarray}
    \mu = \frac{\delta \mathcal F}{\delta \phi},\label{eq:mu}
\end{eqnarray}
where the free energy functional is defined as~\footnote{The functional in eq. \ref{eq:fe} reduces to the functional given in ref.\cite{cates2018theories} under the transformations $\sigma = \sqrt{-8\kappa a/9}$ and $\epsilon = \sqrt{-\kappa/2a}$, after setting $b = -a$ in the form given in the reference. In our formulation, $\sigma$ and $\epsilon$ can be varied independently.}
\begin{eqnarray}
    \mathcal F(\phi, \nabla \phi) = \int d^d\mathbf r \frac{3 \sigma}{4} \left(\frac{1}{4 \epsilon} (\phi^2 - 1)^2 + \epsilon |\nabla \phi|^2\right).\label{eq:fe}
\end{eqnarray}
Here, $\sigma$ represents the interfacial tension, and $\epsilon$ is the width of the diffuse interface between two coexisting phases. For simplicity we assume equal density ($\rho = 1$) and equal kinetic viscosity $\nu$ in both phases. For more general approaches and their comparison see \cite{aland2012benchmark}. Following refs. \cite{padhan2024novel,tiribocchi2015active, das2020transition}, we define the stress tensor $\Sigma(\mathbf r, t)$ as:
\begin{eqnarray}
    \Sigma_{ij} = -\frac{3}{2}\zeta \epsilon \left[(\partial_i\phi) (\partial_j \phi) - \frac{\delta_{ij}}{d} (\nabla \phi)^2\right],\label{eq:stress}
\end{eqnarray}
where the coefficient $\zeta$ is an effective interfacial tension. When $\zeta = 0$, the equations \ref{eq:ch}-\ref{eq:ns} decouple into the Cahn-Hilliard equation (or model B) and the Navier-Stokes equations. According to the active model H~\cite{tiribocchi2015active} the stress becomes extensile for $\zeta > 0$ and contractile for $\zeta < 0$. When $\zeta = \sigma$, these equations reduce to the classical Cahn-Hilliard-Navier-Stokes equations or (passive) model H~\cite{hohenberg1977theory,anderson1998diffuse,kim2004conservative,giorgini2019uniqueness}, where the stress tensor is directly related to the free energy functional through the relation $\nabla \cdot \Sigma = \mu \nabla \phi$ and a modification of $p$, see \cite{Feng_2006}. The deviation $\zeta \neq \sigma$ distinguishes the passive and the active model H. In the active case the stress tensor can not be directly derived from a free energy functional. 

We numerically solve the system of partial differential equations (PDEs) using a Fourier pseudospectral method in a square periodic box similar to the methods used in refs.~\cite{padhan2024novel,perlekar2017two,padhan2023activity}. First, we rewrite eq. \ref{eq:ns} in the vorticity-streamfunction formulation as 
\begin{eqnarray}
    \partial_t \omega + \mathbf{u}\cdot\nabla \omega = \nu \nabla^2\omega + [\nabla \times (\nabla \cdot \Sigma)]\cdot \mathbf{e}_3\label{eq:vs_form},\\
    \omega = (\nabla \times \mathbf{u})\cdot\mathbf{e}_3,\; \omega = - \nabla^{2}\psi,\; \mathbf{u} = \nabla \times (\psi \mathbf{e}_3), \; \nabla \cdot \mathbf{u} = 0,\label{eq:incom}
\end{eqnarray}
where $\mathbf{e}_3$ is the unit vector in $z$-direction and $\omega$ and $\psi$ are vorticity and streamfunction, respectively~\cite{boffetta2012two}. The pressure term is naturally eliminated in this formulation. By utilizing the truncated Fourier series representation for $\phi$,  $\mathbf{u}$, $\omega$ and $\psi$, 
\begin{eqnarray*}
    \phi(\mathbf{r}, t) = \sum_{|k|<N} \hat{\phi}(\mathbf{k}, t) \exp(\dot{\iota} \mathbf{k} \cdot \mathbf{r}),\;\;
    \mathbf{u}(\mathbf{x}, t) = \displaystyle\sum_{|k|<N} \hat{\mathbf{u}}(\mathbf{k}, t) \exp(\dot{\iota} \mathbf{k} \cdot \mathbf{r}), \\
    \omega(\mathbf{r}, t) = \sum_{|k|<N} \hat{\omega}(\mathbf{k}, t) \exp(\dot{\iota} \mathbf{k} \cdot \mathbf{r}),\;\;
    \psi(\mathbf{r}, t) = \sum_{|k|<N} \hat{\psi}(\mathbf{k}, t) \exp(\dot{\iota} \mathbf{k} \cdot \mathbf{r})\,,
\end{eqnarray*}
we write the coupled system of PDEs in Fourier space as follows:
\begin{eqnarray}
    \partial_t \hat{\omega}(\mathbf{k}, t) &=& -\widehat{(\mathbf{u} \cdot \nabla \omega)} (\mathbf{k}, t) - \nu k^2 \hat{\omega}(\mathbf{k}, t) - \dot{\iota} \mathbf{k}\times \widehat{(\nabla \cdot \Sigma)}(\mathbf{k}, t)\,;\label{eq:FFT_omega}\\
    \dot{\iota}\mathbf{k} \cdot \hat{\mathbf{u}}(\mathbf{k}, t) &=& 0\,; \;\;\; \hat{\omega}(\mathbf{k}, t) = \dot{\iota}\mathbf{k}\times \hat{\mathbf{u}}(\mathbf{k}, t)\,;\label{eq:FFT_incom}\\
    \hat{\psi}(\mathbf{k}, t) &=& \frac{\hat{\omega}(\mathbf{k}, t)}{k^2}\,;\label{eq:FFT_stream}\\
    \partial_t \hat{\phi}(\mathbf{k}, t) &=& -\widehat{(\mathbf{u} \cdot \nabla \phi)}(\mathbf{k}, t) - M k^2 \hat{\mu}(\mathbf{k}, t)\,;\label{eq:FFT_phi}\\
    \hat{\mu} (\mathbf{k}, t) &=& \frac{3}{2}\sigma \epsilon k^2 \hat{\phi}(\mathbf{k}, t) + \frac{3}{4}\frac{\sigma}{\epsilon} [\widehat{\phi^3}(\mathbf{k}, t) - \hat{\phi}(\mathbf{k}, t)]\label{eq:FFT_mu}\,, 
\end{eqnarray}
with $\mathbf{k} = k_0\sum_{i=1}^{d} n_i \mathbf{e}_i$, where $k_0 = 2\pi/l$ is the lowest wavenumber and ${n_i}$ are integers with values ranging from $-\frac{N}{2}+1$ to $\frac{N}{2}$; the size of the simulation domain in the real space is $l^d$. We evaluate the nonlinear terms in physical space to avoid convolution and employ the 1/2-dealiasing scheme to eliminate Fourier aliasing errors arising from the Fourier Transforms~\cite{canuto2012spectral, orszag1969numerical, padhan2024novel}. For the time integration we implement the exponential time differencing Runge-Kutta 2 method ETDRK2~\cite{cox2002exponential}. Our program is written in CUDA C and all simulations are carried out in NVIDIA (GPU) A100 processors. The program utilizes the GPU-based Fast Fourier Transform (cuFFT) library. 

For the initial conditions we consider $\omega(\mathbf r, 0) = 0$ and $\phi(\mathbf r, 0) = \phi_0 + \eta(\mathbf r)$, where $\eta(\mathbf r)$ is the noise term distributed uniformly in the interval $[-0.1, 0.1]$. $\phi_0 = 0$ for a symmetric quenching and $\phi_0 \neq 0$ for an asymmetric quenching. Unless otherwise stated, we use a box size of $(2\pi \times 2\pi)$ with $N^2 = 1024^2$ grid points. We fix the following parameters for all simulations: $\epsilon = 0.03,\; M = 5 \times 10^{-4}, \; \sigma = 1.0,\; \nu = 1.5$. We follow \cite{MPCMC_JFM_2013} in choosing $M$, for alternative approaches see \cite{A_ARMA_2009}.
\begin{figure*}
    \centering
    \includegraphics[width=\linewidth]{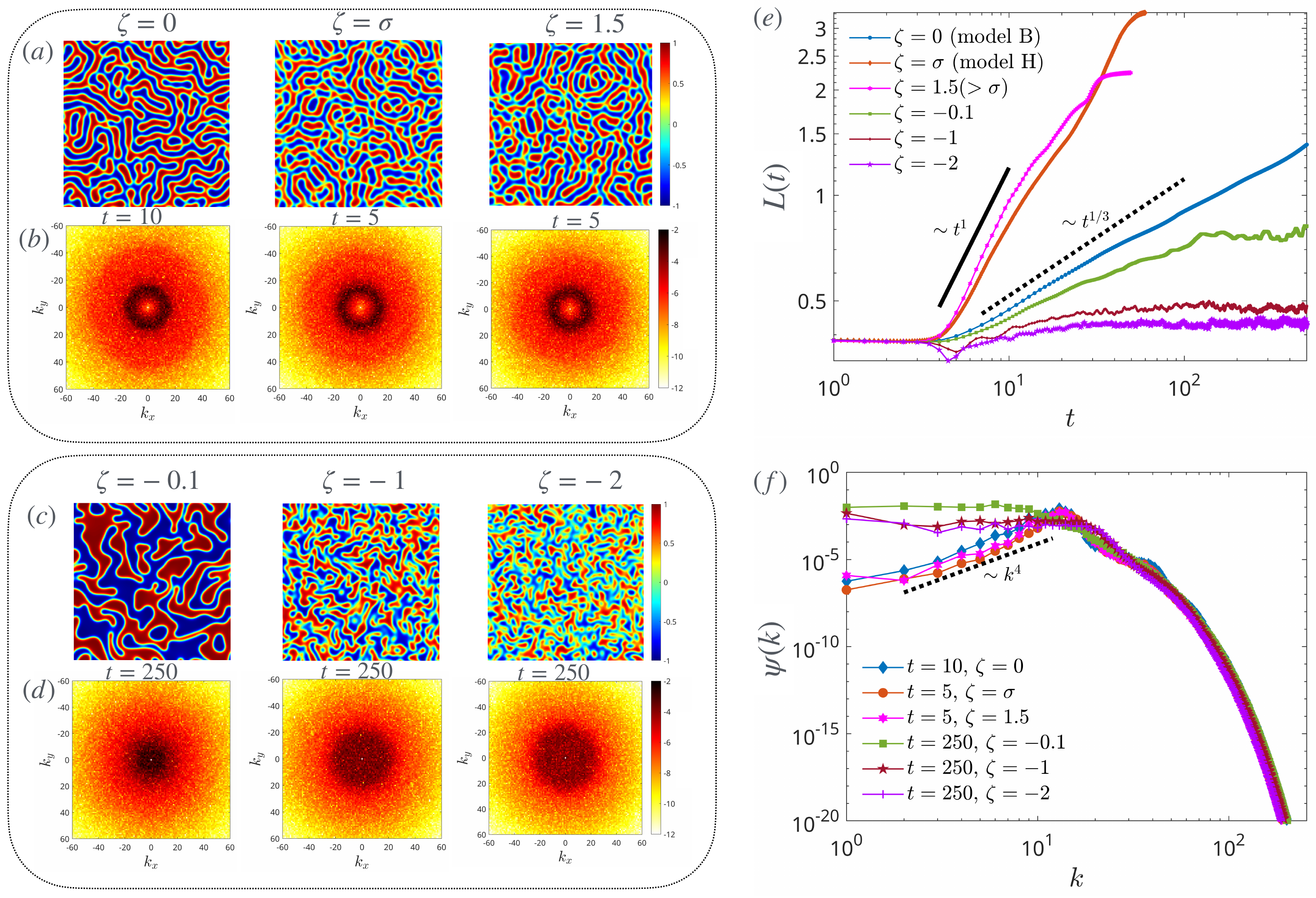}
    \caption{(a) Pseudocolor plots of $\phi$ for $\zeta = 0$, $\sigma$ and $1.5$ taken during the coarsening process, where various scalings have been observed. (b) The corresponding pseudocolor plots of the quantity $|\hat{\phi}(\mathbf{k},t)|^2$ shown in the $(k_x, k_y)$ plane; the colorbars are in logarithmic scale. (c) Pseudocolor plots of $\phi$ for $\zeta = -0.1, \, -1, \, -2$ taken in the statistical steady states after the coarsening process is arrested; (d) the pseudocolor plots of the quantity $|\hat{\phi}(\mathbf{k},t)|^2$ shown in the $(k_x, k_y)$ plane, with the colorbars in logarithmic scale. (e) Log-log of the coarsening length $L(t)$ versus time for various values of $\zeta$. Two scaling regimes have been identified, with scaling exponents $\simeq 1$ and $\simeq 1/3$ for the hydrodynamic and diffusive growth of the domains, respectively. (f) The spectral densities $\psi(k)$ for various values of $\zeta$ at different simulation times. A power-law regime for $k \rightarrow 0$ has been identified with an exponent $\simeq 4$ for $\zeta = 0$, $\sigma$ and $1.5$.}
    \label{fig:all_zeta}
\end{figure*}
\section{Results}
\subsection{Hyperuniformity in passive versus active systems}
To quantify hyperuniformity in $\phi(\mathbf r, t)$ we define the spectral density $\psi(k, t) \equiv S(k,t)/k$, where
\begin{eqnarray}
    S(k, t) = \displaystyle \sum_{k\leq|\mathbf k'|<k+1} |\hat \phi(\mathbf k', t)|^2 
\end{eqnarray}
is the shell-averaged order parameter spectrum for the numerical solution $\hat{\phi}(\mathbf{k}, t)$ of eq. \ref{eq:FFT_phi}~\cite{backofen2024nonequilibrium,perlekar2014spinodal,yadav2024spectral}. Then the coarsening length $L(t)$ follows from $S(k, t)$ as~\cite{perlekar2014spinodal,fan2016cascades}
\begin{eqnarray}
    L(t) = 2\pi \displaystyle\sum_k S(k, t)/\displaystyle \sum_k kS(k, t).
\end{eqnarray}
At late times, $L(t)$ obeys scaling behaviour with $L(t) \sim t^{\lambda}$. Beginning with a symmetric quench ($\phi_0 = 0$), we present the time evolution of $L(t)$ for different values of $\zeta$ in Fig.\ref{fig:all_zeta}(e). When $\zeta = 0$, the coarsening is driven solely by diffusive mechanism as per model B dynamics, and it follows a power-law scaling with an exponent $\lambda \simeq 1/3$, consistent with the classical Lifshitz-Slyozov-Wagner theory~\cite{lifshitz1961kinetics, wagner1961theorie}. For $\zeta = \sigma$, we observe an exponent $\lambda \simeq 1$  as expected from passive model H dynamics~\cite{tiribocchi2015active,perlekar2014spinodal,siggia1979late}. The coarsening process accelerates due to the hydrodynamic effects. In active model H this is even more prominent in the case of extensile stress when $\zeta = 1.5 (> \sigma)$. 
In Fig.\ref{fig:all_zeta}(a), we illustrate the pseudocolor plots of $\phi$ in the scaling regime for $\zeta = 0$, $\sigma$ and $1.5$. The corresponding quantity $|\hat{\phi}(\mathbf{k}, t)|^2$ exhibit ring-like structures as $\mathbf k \rightarrow 0$, a hallmark of hyperuniformity, as shown in Fig.\ref{fig:all_zeta}(b). In active model H dynamics, $\zeta$ can also take negative values for the contractile stress~\cite{tiribocchi2015active}. For $\zeta = -0.1$, $-1$ and $ -2$, coarsening initially sets in but is eventually arrested, leading to non-equilibrium steady states. The initial growth in $L(t)$ is dominated by diffusion, while the coarsening-arrest results from a balance between diffusive growth and anti-growth driven by the contractile stress. The arrest in coarsening occurs earlier for $\zeta = -1$ and $-2$ compared to $-0.1$, which results into a relatively larger domain size in the steady state for $\zeta = -0.1$. This is further illustrated by the pseudocolor plots of $\phi$ at the steady state, shown in Fig.\ref{fig:all_zeta}(c). The corresponding quantity $|\hat{\phi}(\mathbf{k}, t)|^2$ is shown in Fig.\ref{fig:all_zeta}(d), where $|\hat{\phi}(\mathbf{k}, t)|^2 \sim \textit{constant}$ as $\mathbf k \rightarrow 0$. This indicates that the system has lost its hyperuniformity property in the steady state, resulting in non-hyperuniform states. This behaviour is also evident in the 1D spectral density $\psi(k)$, as shown in Fig.\ref{fig:all_zeta}(f). For $\zeta = 0$, $\sigma$ and  $1.5$, we observe the same scaling $\psi(k) \sim k^\alpha$ with $\alpha \simeq 4$. Although the scaling laws for $L(t)$ differ in these cases, hyperuniformity remains preserved. In contrast, for the coarsening arrested systems with $\zeta < 0$ in the steady state, the scaling laws vanish, resulting in $\psi(k) \sim \textit{constant}$. 
\begin{figure*}
    \centering
    \includegraphics[width=\linewidth]{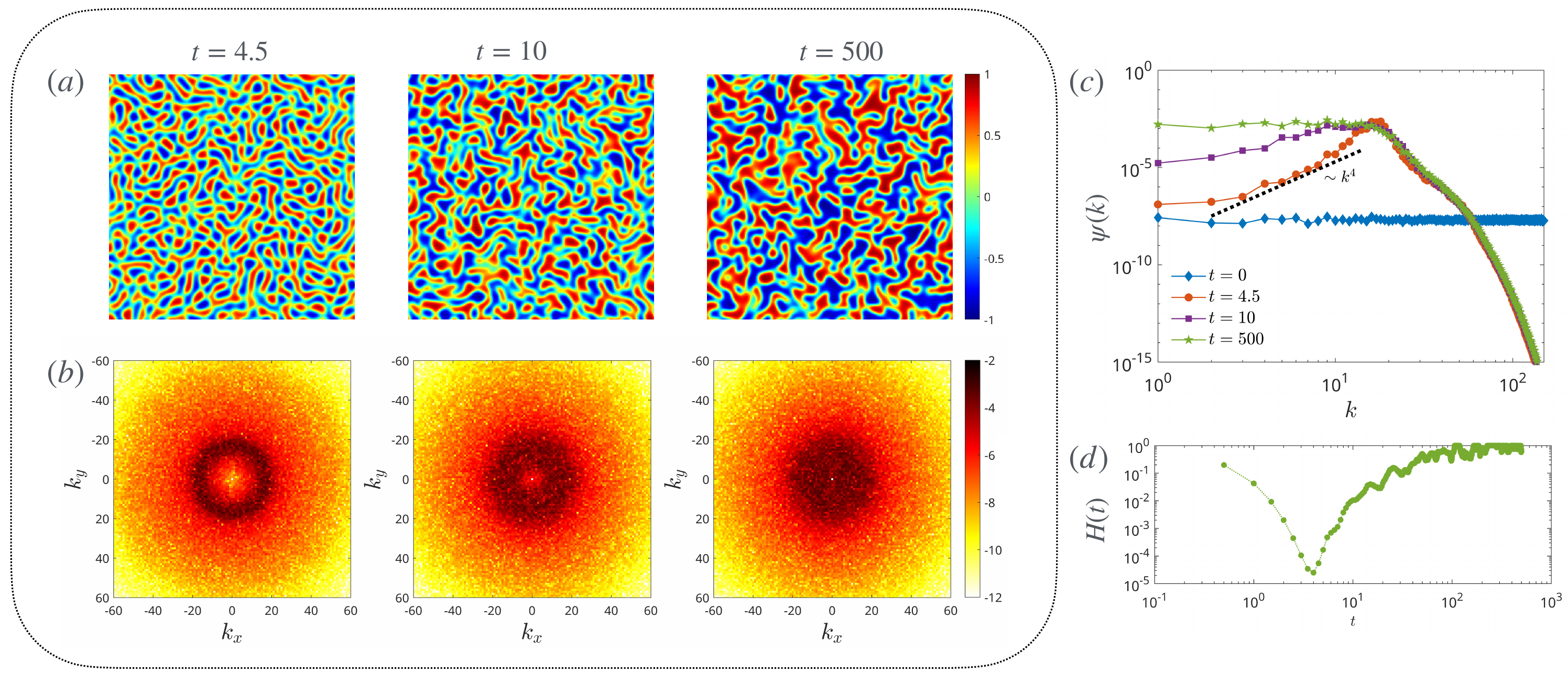}
    \caption{$\zeta = -1$: (a) Pseudocolor plots of $\phi$ at different representative times. (b) The pseudocolor plots of the quantity $|\hat{\phi}(\mathbf{k},t)|^2$ shown in the $(k_x, k_y)$ plane, with the colorbar in logarithmic scale. (c) The 1D spectral densities $\psi(k, t)$ at the corresponding representative times. (d) Logarithmic plot of the hyperuniformity metric $H(t)$.}
    \label{fig:zeta_negative}
\end{figure*}
\subsection{Hyperuniformity to non-hyperuniformity transition}
We observe a smooth transition from hyperuniformity to non-hyperuniformity for negative $\zeta$, where the coarsening is arrested. This transition occurs during the coarsening process, before reaching the steady state. We illustrate this transition for $\zeta = -1$ in Fig.\ref{fig:zeta_negative} using the pseudocolor plots of $\phi(\mathbf{r}, t)$ and $|\hat{\phi}(\mathbf{k}, t)|^2$, along with the spectral density $\psi(k, t)$ and hyperuniform parameter $H(t)$, which is defined below. For the initial condition for $\phi$ the spectral density is uniform across all wave numbers. As time progresses, the spectral density exhibit hyperuniformity with $\alpha \simeq 4$ at time $t = 4.5$ (orange line in Fig.\ref{fig:zeta_negative}(c)), similar to the passive phase separation discussed in the previous section. However the scaling exponent changes over time, giving less steep power-law exponents $\alpha < 4$ at $t = 10$ (purple line). Eventually, the system reaches to a non-hyperuniform state with $\psi(k) \sim \textit{constant}$ (green line), which acts as a coloured noise with a cut-off frequency related to the steady state length scale. We quantify the degree of hyperuniformity using the following hyperuniformity metric~\cite{ma2017random,backofen2024nonequilibrium}
\begin{eqnarray}
    H(t) = \frac{\psi(1, t)}{\psi(k_{peak}, t)},
\end{eqnarray}
where $k_{peak}$ corresponds to the peak of the spectrum shown in Fig.\ref{fig:zeta_negative}(c). For an ideal hyperuniform system we have $H = 0$. However, $H \leq 10^{-4}$ is typically considered as effectively hyperuniform and $H \leq 10^{-2}$ as nearly hyperuniform~\cite{ma2017random}. We compute $H$ at each time instance and show the evolution $H(t)$ in Fig.\ref{fig:zeta_negative}(b). During the coarsening process, $H(t)$ reaches its minimum value $\simeq 10^{-5}$, where the system shows effective hyperuniformity with $\alpha \simeq 4$. At the intermediate times, the system shows nearly hyperuniformity with $10^{-4}<H<10^{-2}$ and $\alpha < 4$. Finally, the system make a transition to a non-hyperuniform state with $H \simeq 1$. During such transitions, the morphologies in $\phi$ also change, as shown at various representative times in Fig.\ref{fig:zeta_negative}(a). The quantity $|\hat{\phi}(\mathbf{k}, t)|^2$ likewise evolve for $\mathbf k \rightarrow 0$, shifting from ring to uniform structures, as illustrated in Fig.\ref{fig:zeta_negative}(b).

\subsection{Transfer mechanism for hyperuniformity and non-hyperuniformity}
\begin{figure*}
    \centering
    \includegraphics[width=\linewidth]{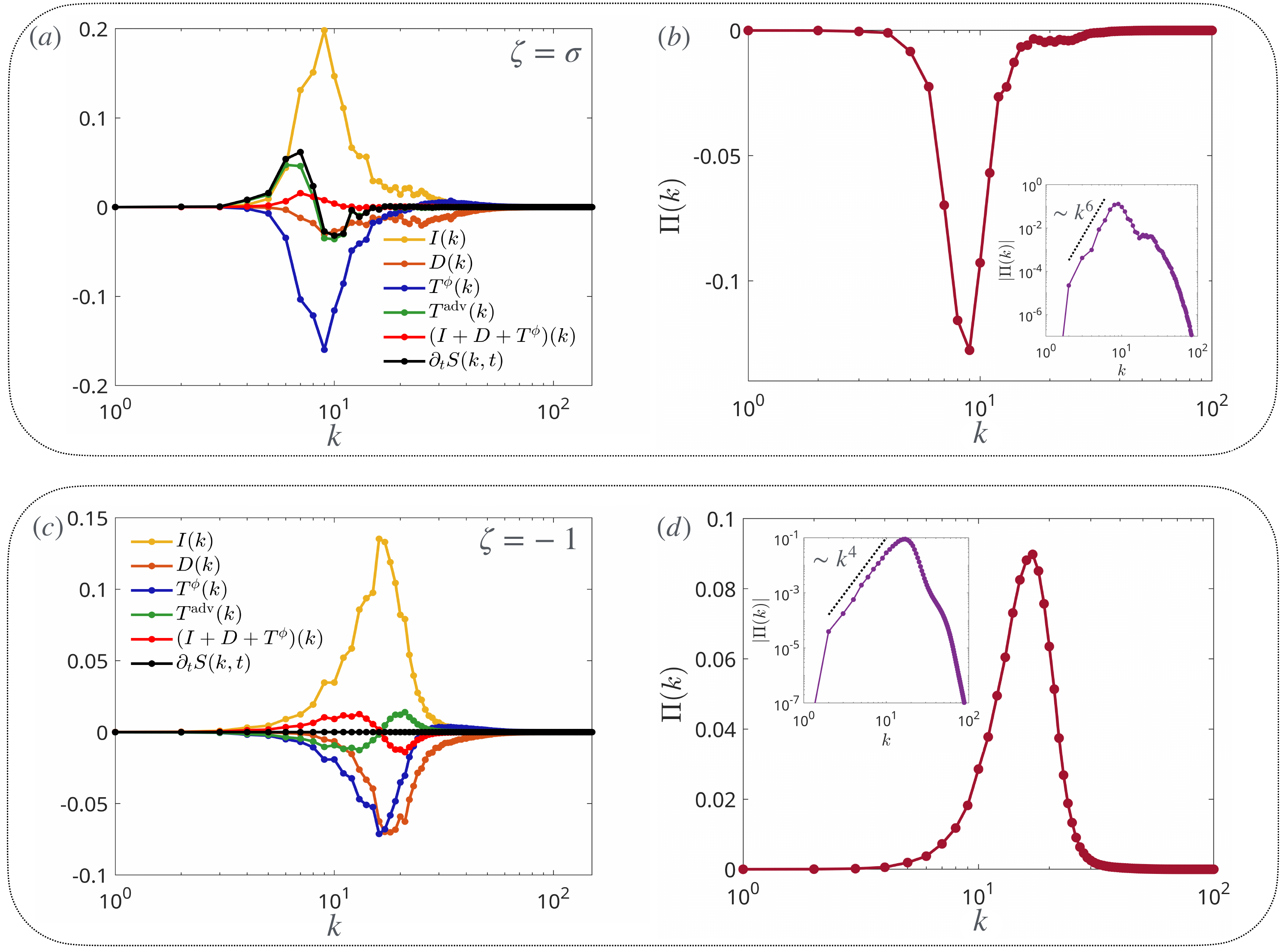}
    \caption{(a) Shell-averaged spectral forms of different contributions in Eq.\ref{eq:budget} for $\zeta = \sigma$ at $t = 8$, during the regime where $L(t) \sim t^1$. (b) Plot versus $k$ of the flux associated with the advection term. The horizontal axis is in the logarithmic scale. The inset shows the absolute value of the flux in the logarithmic scale. The $k\rightarrow 0$ region is fitted to a power-law with exponent $\simeq 6$. (c) Spectral forms of different contributions in Eq.\ref{eq:budget} for $\zeta = -1$. The spectra are obtained by averaging $700$ datasets in the steady state. (d) Plot of the flux in wavenumber space, with the inset showing the log-log plot of its absolute value. A power-law scaling of $\sim k^4$ is observed in the small $k$ regime.}
    \label{fig:flux}
\end{figure*}

We now present a scale-by-scale analysis by using the order parameter spectrum equation to investigate the  mechanisms of spectral energy or structural energy transfer. This approach is analogous to the methods used to study scale-by-scale energy transfer mechanism through kinetic energy budget equation in fluid turbulence literature~\cite{verma2019energy,alexakis2018cascades, carenza2020cascade}. There are also several such analysis on both passive and active model H dynamics, with a focus on the fluid's kinetic energy~\cite{perlekar2014spinodal,perlekar2017two, perlekar2019kinetic,padhan2024novel}. However, similar analysis for order parameter $\phi$ has not been attempted so far in such models. Recently, this approach has also been applied to study domain growth in the Cahn-Hilliard equation~\cite{yadav2024spectral}. We derive the spectral budget equation for $S(k, t)$ from eqs.\ref{eq:FFT_phi} and \ref{eq:FFT_mu}, using the relation $\partial_t |\hat \phi(\mathbf{k}, t)|^2 = \partial_t \hat\phi(\mathbf{k}, t) \hat\phi(\mathbf{-k}, t) + \partial_t \hat \phi(\mathbf{-k}, t) \hat \phi(\mathbf{k}, t)$, expressed in terms of shell-averaged variables:
\begin{eqnarray}
    \partial_t S(k, t) = I(k, t) + D(k, t) + T^{\phi} (k, t) + T^{\mathrm{adv}}(k, t)\label{eq:budget},
\end{eqnarray}
where
\begin{eqnarray}
    I(k, t) &=& \frac{3\sigma}{2\epsilon} M k^2 S(k, t)\\
    D(k, t) &=& -3M\sigma \epsilon k^4 S(k, t)\\
    T^{\phi}(k, t) &=& -\frac{3\sigma}{2\epsilon} M k^2 \;\displaystyle \mathlarger{\mathfrak {R}}\left[\displaystyle \sum_{k \leq |\mathbf{k'}| < k+1} \widehat{\phi^3}(\mathbf{k'}, t) \hat{\phi}(\mathbf{-k'}, t)\right]\\
    T^{adv}(k, t) &=& -\mathlarger{\mathfrak R}\left[\displaystyle \sum_{k \leq |\mathbf{k'}| < k+1} \widehat{(\mathbf u \cdot \nabla \phi)}(\mathbf{k'}, t) \hat \phi(\mathbf{-k'}, t)\right].
\end{eqnarray}
The terms $I(k, t)$ and $D(k, t)$ are the rate of (structural) energy injection and dissipation, respectively. $T^{\phi}(k, t)$ and $T^{adv}(k, t)$ are the nonlinear contributions associated with the cubic nonlinearity and advection term, respectively. In Fig.\ref{fig:flux}(a) and (c), we illustrate all the terms in eq.\ref{eq:budget} for the cases $\zeta = \sigma$ (passive model H, leading to hyperuniformity) and $\zeta = -1$ (active model H, leading to non-hyperuniformity), respectively. For $\zeta = \sigma$, we show the spectral contributions at time $t = 8$ in the scaling regime $L(t) \sim t^1$. But, for $\zeta = -1$, we present the time-averaged spectra in the steady state. For both these cases, $I(k) \geq 0$ (yellow line) and thus injection of structural energy into the system, while $D(k)$ (orange line) indicates dissipation of energy at large $k$. The cubic nonlinear contribution $T^{\phi}(k)$ (blue line) acts as an energy sink and drives domain formation in the system. The combined term $(I+D+T^{\phi})(k) \equiv I(k) + D(k) + T^{\phi}(k)$ (red line) remains positive as $k \rightarrow 0$ and thus governs coarsening in the system. In passive Cahn-Hilliard dynamics, this term drives the system towards hyperuniformity~\cite{de2024hyperuniformity}. The contribution of the advection term then determines whether hyperuniformity is preserved. As we explain below, $T^{adv}(k)$ retains hyperuniformity when $\zeta = \sigma$ (passive model H) and suppresses it when $\zeta = -1$ (active model H). In Fig.\ref{fig:flux}(a), we observe $\partial_t S(k, t) \geq 0$ for $k \rightarrow 0$, which indicates coarsening in the system. This coarsening is dominated by the hydrodynamics as $T^{adv}(k) \gg (I+D+T^{\phi})(k)$, which explains the faster domain growth $L(t) \sim t^1$ compared to diffusive growth $L(t) \sim t^{1/3}$. On the other hand in Fig.\ref{fig:flux}(b), $\partial_t S(k, t) \simeq 0$ resulting in $|T^{adv}(k)| \simeq |(I+D+T^{\phi})(k)|$. This indicates that coarsening is arrested in the steady state due to a balance between diffusive growth and hydrodynamic anti-growth for $k \rightarrow 0$. Theoretically, this is consistent with $\partial_t S(k, t) = 0$ in the steady state, which results in $T^{adv}(k) = - (I+D+T^{\phi}(k)$. We now describe the transfer mechanism underlying hyperuniformity and non-hyperuniformity, which is connected to the evolving domain structures in the phase-separating system. This mechanism explains how smaller structures grow into larger ones or, conversely, how larger structures break down into smaller ones.
Using the incompressible condition $\nabla \cdot \mathbf{u} = 0$ in eq. \ref{eq:ns} we obtain $\int_{0}^{\infty} T^{adv}(k) dk = 0$, meaning the advection term acts neither as a source nor a sink. This implies that the total (structural) energy transfer across all wavenumbers sums to zero, which is why the area under the curves $T^{adv}(k)$ (green lines) in Fig.\ref{fig:flux}(a),(c) is zero. The sole role of this term is to redistribute the energy across scales. So, the term $T^{adv}(k)$ describes the local rate of energy transfer or cascade at wavenumber $k$. For $\zeta = \sigma$,  $T^{adv}(k)$ transfers energy from large $k$ to small $k$, resulting in an inverse flow of the energy in $k$-space. Conversely, for $\zeta = -1$, $T^{adv}(k)$ moves energy from small $k$ to large $k$, yielding a forward flow of the energy. This transfer mechanism is further quantified using the flux $\Pi(k)$ (see eq. \ref{eq:flx}), which is defined as the net rate at which energy is transferred across a given wavenumber $k$ due to interactions at neighbouring $k$~\cite{alexakis2018cascades,verma2019energy,perlekar2019kinetic}.
\begin{eqnarray}
    \Pi(k) = -\int_{0}^{k} T^{adv}(k') dk'\label{eq:flx}
\end{eqnarray}
The negative flux in Fig.\ref{fig:flux}(b), shown for $\zeta = \sigma$, indicates an inverse cascade present in the hyperuniform system. The log-log plot of absolute flux $|\Pi(k)|$ suggests a power-law scaling with an exponent $\simeq 6$. Due to the steepness of the power-law, we can consider $|\Pi(k)| \rightarrow 0$ as $k \rightarrow 0$. It suggests that the flux is negligible at small $k$ and only effective at relatively large $k$. So the power law scaling that we observe in $\psi(k)$ remains unchanged for $k \rightarrow 0$, even with the presence of hydrodynamic effects. In contrast, we observe the opposite behaviour for $\zeta = -1$ as shown in Fig.\ref{fig:flux}(d). Here, the flux is positive, indicating a forward cascade associated with the non-hyperuniform behaviour in the active system. The flux also follows a power-law with a scaling exponent $\simeq 4$. This forward cascade causes the scaling exponent in $\psi(k)$ to becomes less steep over time ($\alpha < 4$) (see Fig.\ref{fig:zeta_negative}). This means that the system progressively looses its hyperuniformity characteristics. Eventually the system reaches a non-hyperuniform state, where $\psi(k) \sim \textit{constant}$ and the system remains in the steady state as per the law of equipartition of energy. 

\subsection{Turbulence in hyperuniform and non-hyperuniform systems}
\begin{figure*}
    \centering
    \includegraphics[width=\linewidth]{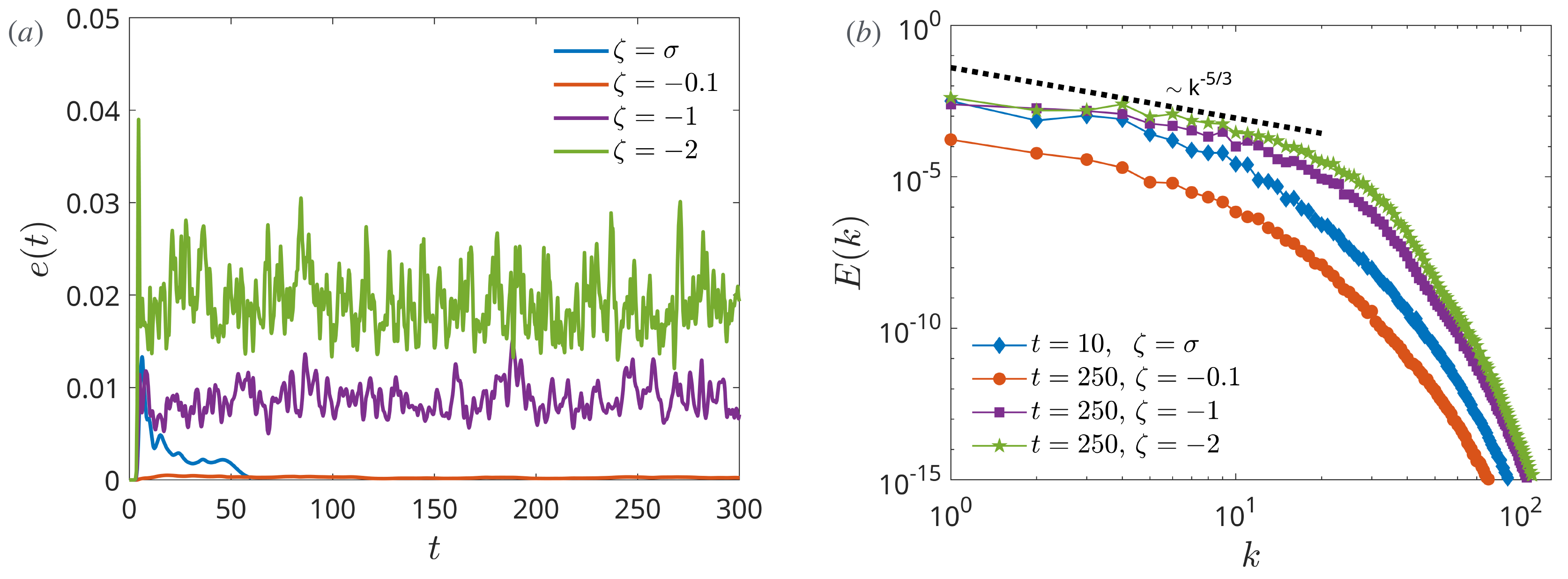}
    \caption{(a) The plot of kinetic time series for $\zeta=\sigma, 0.1, -1, -2$. (b) Log-log plots of the kinetic energy spectrum for various values of $\zeta$. The dotted black line represents a fit with $\sim k^{5/3}$.}
    \label{fig:en_spec}
\end{figure*}
The flow is entirely induced by the stress term in eq. \ref{eq:ns}, as there is no external forcing present in the system. This flow field, in turn, influences the dynamics of $\phi$ through the advection term. As described in the previous section, this term drives the transition of the system from hyperuniform to non-hyperuniform states. Here, we discuss the properties of the flow field in such states. Fig.\ref{fig:en_spec}(a) shows the kinetic energy time series $e(t)$ for $\zeta = \sigma$, $-0.1$, $-1$ and $-2$, where $e(t) = \sum_k E(k, t)$, and the kinetic energy spectrum $E(k, t)$ is defined as
\begin{eqnarray}
    E(k, t) = \displaystyle \sum_{k\leq|\mathbf k'|<k+1} |\hat {\mathbf u}(\mathbf k', t)|^2.
\end{eqnarray}
For the systems undergoing complete phase separation ($\zeta > 0$), the kinetic energy approaches zero at the long time limit, where the systems reaches equilibrium. In contrast, for the non-equilibrium steady states ($\zeta < 0$) the kinetic energy remains non-zero, reflecting the persistence of non-equilibrium steady flow states. The flow field exhibits a broad kinetic energy spectrum, a signature of turbulence. These spectra, shown for various $\zeta$ in Fig.\ref{fig:en_spec}(b), follow a power-law scaling with an exponent of  $\simeq -5/3.$ 
Interestingly, both hyperuniform and non-hyperuniform states show similar turbulent behaviours. The flow can be characterized by the Reynolds number, $\textrm{Re} = U_{rms}L_I/\nu$, where $U_{rms} = \sqrt{\sum_k E(k)}$ is the root-mean-squared velocity, and $L_I$ is the integral length scale defined as $L_I = 2\pi \sum_k E(k)/\sum_k k E(k)$. For the parameters used in our simulations, Re is very small. For instance, for $\zeta = -1$, $\textrm{Re} = 0.1$. This low-Reynolds-number turbulence arises solely from the stress induced by $\phi$, in the absence of external energy sources or an inertial energy cascade. A detailed analysis of such active-scalar turbulence, including scale-by-scale analysis of $E(k, t)$, is provided in ref.~\cite{padhan2024novel}. In contrast to hyperuniformity in $\phi$ we now examine hyperuniformity in the vorticity field. By defining the (vorticity) spectral density $\psi_{\omega}(k, t)$, which is associated with $\hat{\omega}(\mathbf{k}, t)$ by  $\psi_{\omega}(k, t) \equiv k E(k, t)$~\cite{backofen2024nonequilibrium}, it follows that $\psi_{\omega}(k) \sim k^{-2/3}$, which suggests anti-hyperuniformity of the vorticity. 

\begin{figure*}
    \centering
    \includegraphics[width=\linewidth]{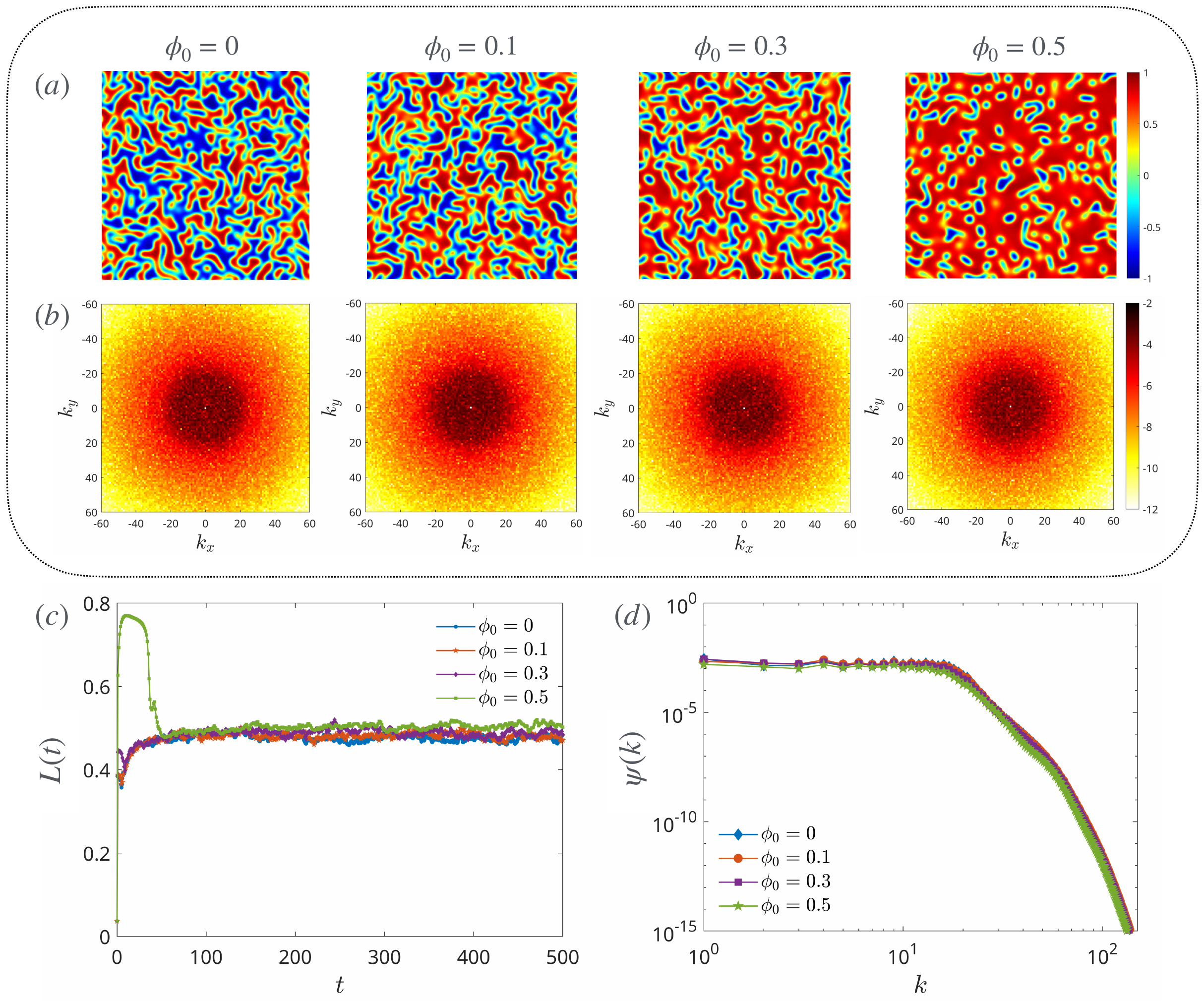}
    \caption{$\zeta = -1$: (a) Pseudocolor plots of $\phi$ for different value of $\phi_0$ at the same represtative times in the steady state. (b) The corresponding pseudocolor plots of $|\hat{\phi(\mathbf{k}, t)}|^2$ shown in the $(k_x, k_y)$ plane, with the colorbar in logarithmic scale. (c) The time evolution of the coarsening length $L(t)$ for different values of $\phi_0$. It highlights the transition to statistical steady states.(d) The time averaged 1D plots of the spectral densities $\psi(k)$.}
    \label{fig:density}
\end{figure*}
\subsection{Impact of asymmetric quench and finite-size effects on the hyperuniformity} 
We now investigate whether the loss of hyperuniformity, as discussed in the previous sections and shown in Fig.\ref{fig:zeta_negative}, depends on the nature of the quench implemented through the initial condition. So far, we have described the results for a symmetric quench ($\phi_0 = 0$). We now explore this question for asymmetric quenches with $\phi_0 \neq 0$ and $\phi_0 < 1/\sqrt 3$, where $\phi_0 = 1/\sqrt 3$ is the spinodal point for the chosen double-well potential in the free energy functional given in eq. \ref{eq:fe}~\cite{cates2018theories}. Keeping the activity $\zeta = -1$ and other parameters fixed, we vary $\phi_0$. The system reaches non-equilibrium steady states for all values of $\phi_0$ as shown in the plot of $L(t)$ in Fig.\ref{fig:density}(c). We present the pseudocolor plots of $\phi$ and  $|\hat{\phi}(\mathbf{k},t)|^2$ in Fig.\ref{fig:density}(a)-(b) at the non-equilibrium steady states. Interestingly, the spectral densities are similar across different values $\phi_0$, despite differences in the morphologies of $\phi$. Specifically, we observe droplet formation for $\phi_0 = 0.5$. We conclude that the system shows non-hyperuniform behaviour independent of the values of $\phi_0$, which is also reflected in the spectrum as shown in Fig.\ref{fig:density}(d), where $\psi(k) \sim \textit{constant}$. 
\begin{figure*}
    \centering
    \includegraphics[width=\linewidth]{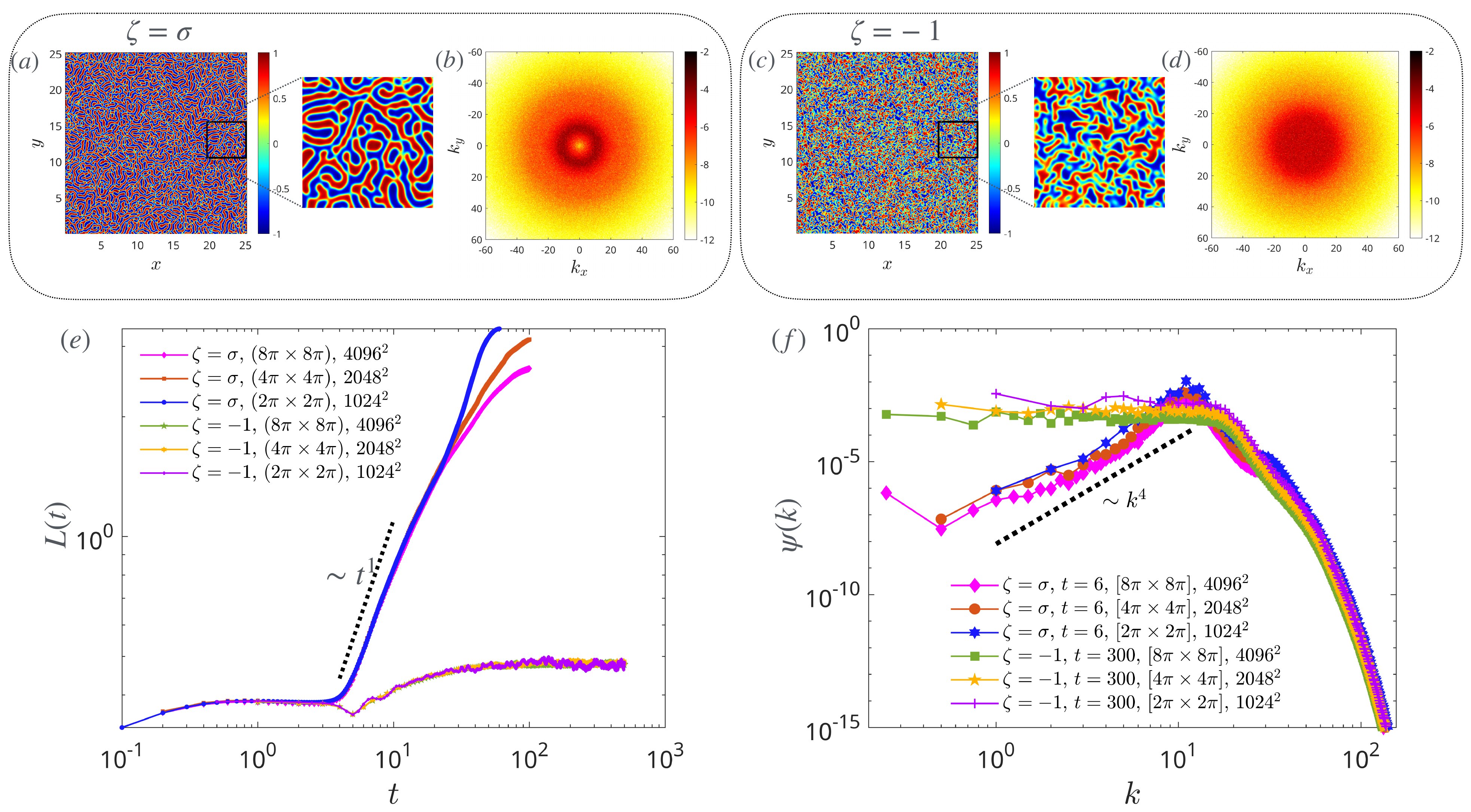}
    \caption{(a) Pseudocolor plot of $\phi$ for $\zeta = \sigma$ at time $t = 6$. The size of the simulation domain is $(8\pi, 8\pi)$ with $4096^2$ collocation points. A zoom-in plot is shown for the area bounded by the black box. (b) The corresponding pseudocolor plot of $|\hat{\phi}(\mathbf{k}, t)|^2$ shown in the $(k_x, k_y)$ plane; the colorbar is shown in the logarithmic scale for better visulization. (c) The pseudocolor plot of $\phi$ for the case $\zeta = -1$ at a representative time in the steady state. (d) The corresponding pseudocolor plot of $|\hat{\phi}(\mathbf{k}, t)|^2$ shown in the $(k_x, k_y)$ plane. (e) Log-log plot of the coarsening length $L(t)$ versus time for $\zeta = \sigma, -1$ taken from simulations with different box size and number of grid points. (f) The 1D plots of the spectral densities for $\zeta = \sigma, -1$ for various simulation domains.}
    \label{fig:box_size}
\end{figure*}

To verify that the observations made in this paper (obtained for a box size of $2\pi\times2\pi$ and $1024^2$ grid points) are not influenced by the finite-size effects, we perform numerical simulations in larger domains by keeping other simulation parameters fixed. The scaling laws for $L(t)$ and $\psi(k)$ remain unchanged across different box sizes, as shown in Fig.\ref{fig:box_size}(e)-(f) for $4\pi\times 4\pi$ and $8\pi\times 8\pi$. Additionally, we present the pseudocolor plots of $\phi(\mathbf{r},t)$ and $|\hat{\phi}(\mathbf{k},t)|^2$ at representative times for $\zeta = \sigma$ and $\zeta = -1$ in $8\pi\times 8\pi$ domain, see Fig.\ref{fig:box_size}(a)-(d).

\section{Conclusion}
We have demonstrated, through direct numerical simulations, the significant role of hydrodynamic effects in influencing hyperuniformity in scalar field theories. Our findings reveal that in passive phase separation, governed by the Cahn-Hilliard-Navier-Stokes equations (passive model H), the system exhibits hyperuniform behavior, akin to that observed in the Cahn-Hilliard equation (passive model B). However, in contrast to active model B or active model B+, which also show hyperuniformity, we have observed a transition from hyperuniformity to non-hyperuniformity in active phase separation governed by active model H considering contractile active stress. Through a scale-by-scale analysis of the spectral equations for the order parameter, we have demonstrated that hyperuniformity and non-hyperuniformity are determined by the nature of spectral energy transfer in these systems. Hyperuniformity arises from an inverse cascade, which promotes large-scale pattern formation through domain growth. In contrast, non-hyperuniformity results from a forward cascade, where large-scale structures continuously break down into smaller ones. Moreover, our study reveals that the transition between hyperuniformity and non-hyperuniformity is independent of the quenching approach and the size of the simulation domain.\\

\noindent
{\bf{Acknowledgements:}} We acknowledge computing resources at FZ Jülich
under grant MORPHO and at ZIH under grant WIR. We further acknowledge fruitful discussions with R. Backofen.

\section*{References}
\bibliographystyle{unsrt}
\bibliography{main}
\end{document}